# Intrinsic layer polarization and multi-flatband transport in non-centrosymmetric mixed-stacked multilayer graphene

Kai Liu[1,2†], Yating Sha[1,2†], Bo Yin[3,4†], Hongyun Zhang[5,6], Jinxi Lu[5], Shuhan Liu[2], Size Wu[1,2], Yulu Ren[2], Zhongxun Guo[7], Jingjing Gao[7], Ming Tian[8], Neng Wan[7,8], Kenji Watanabe[9], Takashi Taniguchi[10], Bingbing Tong[3,11], Guangtong Liu[3,11], Li Lu[3,11], Yuanbo Zhang[7], Weidong Luo[2], Zhiwen Shi[2], Shuyun Zhou[5,12], Quansheng Wu[3,4], Guorui Chen[1,2*]

[1]*Tsung-Dao Lee Institute, Shanghai Jiao Tong University, Shanghai, 201210, China*

[2]*Key Laboratory of Artificial Structures and Quantum Control (Ministry of Education), School of Physics and Astronomy, Shanghai Jiao Tong University, Shanghai 200240, China*

[3]*Beijing National Laboratory for Condensed Matter Physics, and Institute of Physics, Chinese Academy of Sciences, Beijing 100190, China*

[4]*University of Chinese Academy of Sciences, Beijing 100049, China*

[5]*State Key Laboratory of Low-Dimensional Quantum Physics and Department of Physics, Tsinghua University, Beijing 100084, P. R. China*

[6]*Advanced Institute for Materials Research (WPI-AIMR), Tohoku University, Sendai 980-8577, Japan*

[7]*State Key Laboratory of Surface Physics and Department of Physics, Fudan University, Shanghai 200433, China*

[8]*Key Laboratory of MEMS of Ministry of Education, School of Integrated Circuits, Southeast University, Jiangsu, Nanjing 210096, China*

[9]*Research Center for Electronic and Optical Materials, National Institute for Materials Science, 1-1 Namiki, Tsukuba, Japan*

[10]*Research Center for Materials Nanoarchitectonics, National Institute for Materials Science, 1-1 Namiki, Tsukuba, Japan*

[11]*Hefei National Laboratory, Hefei, Anhui 230088, China*

[12]*Frontier Science Center for Quantum Information, Beijing 100084, P. R. China*

[†]These authors contributed equally to this work.

*Correspondence to: chenguorui@sjtu.edu.cn

**Abstract:** Graphene multilayers exhibit electronic spectra that depend sensitively on both the number of layers and their stacking order. Beyond trilayer graphene, mixed stacking sequences (alternating Bernal and rhombohedral layers) give rise to multiple coexisting low-energy bands. Here we investigate ABCBC-stacked pentalayer graphene, a less-studied non-centrosymmetric mixed sequence. This stacking can be regarded as an ABC (rhombohedral) trilayer on top of an AB (Bernal) bilayer, so its low-energy band structure contains both a cubic band and a parabolic band that hybridize. In transport measurements, we observe an intrinsic band gap at charge neutrality whose magnitude changes asymmetrically under an applied perpendicular displacement field. This behavior reflects the spontaneous layer polarization inherent to the broken inversion symmetry and mirror symmetry. By tuning the displacement field and carrier density, we drive multiple Lifshitz transitions in the Fermi surface topology and realize Landau levels with different degeneracies arising from the multi-flatband system. Remarkably, a $\nu$ = -6 quantum Hall state emerges at an exceptionally low magnetic field (~20 mT), indicating the interplay between spontaneous symmetry breaking and Berry curvatures. Our results establish mixed-stacked multilayer graphene as a tunable platform with various broken symmetries and multiple flatbands, suitable for exploring emergent correlated electronic states.

**Main text:**

Graphene multilayers present a versatile two-dimensional platform because their low-energy electronic spectra depend strongly on both layer number and stacking order. The simplest cases are Bernal (AB) stacking and rhombohedral (ABC) stacking, which yield quite different quasiparticle dispersions. For example, Bernal-stacked trilayer (ABA) graphene hosts both massless and massive carriers, whereas rhombohedral-stacked trilayer (ABC) hosts chiral quasi-flat bands[1–6]. Above trilayer, mixed stacking sequences emerge by alternating AB and ABC layers, greatly enriching the possible electronic structures[7–11]. For instance, there are four possible mixed stackings in pentalayer graphene (see Fig. la) that are not equivalent with each other.

Mixed-stacked graphene could be regarded as the natural crystal hosts multi-flatbands with special symmetries[7]. When considering the electronic band structures, a better view of multilayer graphene is a decomposition of chiral layers (i.e. R-layers), in which the band structures follow "partitioning rules"[3]. Following this rule, we can quickly capture the simplified band structures of graphene multilayers when only considering nearest interlayer and intralayer hopping terms. Dashed lines in Fig. 1a decomposes all possible pentalayer graphene into chiral layers, where ABCBC is decomposed into a chiral trilayer and a chiral bilayer (3+2). As a result, ABCBC's simplified band structure is a combination of an ABC cubic band and an AB parabolic band, which are spatially located in corresponding chiral layers (Fig. 2d). At the same time, M-stacked graphene can host special lattice symmetries other than B- or R-stacking. Colored panels in Fig. 1a shows three symmetry groups of all six stackings in pentalayer: ABABA(B) and ABCBA have mirror symmetry, ABCAB(R) and ABACA have inversion symmetry, and ABCBC and ABCAC have non-centrosymmetry. Therefore, from the electronic and lattice structures, M-stacked graphene is an interesting natural crystal system to be explored.

Experimentally, mixed-stacked graphene is less studied because of the lack of efficient way to identify the stacking order and its fragility to slide stacking sequence. Recent advances in scanning near-field optical microscope (SNOM)[12–14] and low-stress van der Waals transfer technique now allow direct imaging of stacking domains and fabricate high-quality metastable stacking devices[12–17]. We use SNOM to locate mixed-stacked (M-type) regions in pentalayer graphene: in Fig. 1b the SNOM contrast clearly distinguishes Bernal (intermediate signal), rhombohedral (dark), and mixed (bright) domains. We find mixed domains are generally smaller and appear in roughly 30% of flakes. Once identified, these regions are isolated by AFM cutting[18], encapsulated in hBN and processed into dual-gated devices. We verified if the stacking is preserved after hBN encapsulation by performing phonon-polariton assisted SNOM[14] through the heterostructure (Fig. S1). In total we fabricated four ABCBC devices (labeled M1–M4), which show consistent transport behavior (data below are from M1 unless noted). The NanoARPES measurement of device M4 (Fig. 1c and Fig. S2) proves its electronic dispersion satisfies the single particle band structure of ABCBC (the NanoARPES image shows two bands at low-energy level, and other three isolated bands at high-energy level, being well fitted by the calculated band structure).

In this work, we focus on the transport signatures of the non-centrosymmetric ABCBC stacking. Our measurements reveal an intrinsic band gap at charge neutrality even at zero applied field, and this gap responds asymmetrically to a vertical displacement field. We also observe a rich Landau level diagram when a magnetic field is applied, including regions of unusual Landau level degeneracies and multiple Lifshitz transitions as the Fermi energy moves through the flat bands. Notably, a quantum Hall state at filling $\nu$ = -6 develops at a remarkably low magnetic field of ~20 mT. These findings demonstrate that mixed-stacked pentalayer graphene naturally realizes multiple tunable flat bands and broken inversion symmetry, making it an attractive platform for exploring emergent many-body and topological states.

**Intrinsic band gap from non-centrosymmetric stacking**

Figure 2a shows the measured resistance $R_{xx}$ of the ABCBC device (M1) as functions of $n$ and $D$. Strikingly, the data exhibit a pronounced asymmetry with $D$ around charge neutrality. For centrosymmetric stackings (ABA or ABC), one expects $R_{xx}$ to be symmetric between $D$ and $-D$[19,5,20–22,6,23], due to mirror or inversion symmetry. Here, however, the resistance at $n = 0$ is high for negative $D$ and low for positive $D$, indicating a built-in polarization.

To quantify this, for $n = 0$, we measure the temperature dependence of $R_{xx}$ at different $D$ to extract the activation gap (Fig. 2b and Fig. S4). We find a finite gap of about 0.65 meV at $n = D = 0$. As $D$ is made negative, the gap first increases then decreases; while for positive $D$ it rapidly closes to zero. This non-monotonic, asymmetric gap versus $D$ is in good agreement with our band structure calculations for ABCBC (see methods and Fig. S10). The key point is that even at $D = 0$ there is an intrinsic gap: the ABCBC stacking itself lacks inversion symmetry, so the atomic sites in different layers feel different chemical environments and therefore host different onsite energies. This results in opposite built-in electric fields and gaps in the trilayer (ABC) and bilayer (AB) blocks. Considering the interlayer interactions (interlayer hopping) between trilayer and bilayer, the band hybridization leaves a small net gap (Fig. 2d(ii)). When $|D| \neq 0$, as shown in Fig. 2d(i) and (iii), since the total displacement field for each chiral layer is the sum of built-in and external one, the gaps of cubic and parabolic bands are different. At the same time, the sub-bands shift in energy in an opposite direction under external $D$, so the gap closes at $+|D|$ and shows nonmonotonic behavior at $-|D|$.

This picture is further supported by magneto transport at neutrality. For $D > 0$, the conduction band of the bilayer overlaps the valence band of the trilayer (Fig. 2d(iii)), so electrons and holes coexist. Indeed, in Fig. 2c, $R_{xx}$ grows approximately as $B^2$, and the magnetoresistance reaches about 50,000% at 6 T, a signature of two-carrier transport[24,25]. By contrast, for $D < 0$, the bands remain separated and no linear- $B^2$ magnetoresistance is seen. Thus, the transport data confirm that the ABCBC stacking has an intrinsic layer polarization and band gap, tunable by the external displacement field. Similar transport behaviors are reproduced in device M2–M4 (see Fig. S3, 5, 6).

**Tunable Lifshitz transitions**

When the Fermi level is moved away from neutrality, the multi-band nature of ABCBC leads to complex transport features. At a small magnetic field ($B$ = 1 T), the Landau levels (LLs) in Fig. 3a reveal multiple sequences of quantum oscillations. By analyzing the LL degeneracies, we identify several distinct regions shown in Fig. 3b. Close to neutrality (region I) the resistivity is high. Regions labeled II show LLs of degeneracy 4 (spin and valley degenerate), corresponding to a single ordinary Fermi pocket in each band. At positive $D$ and hole-doped region labeled IV, LL degeneracy is 12, implying three coexisting pockets from triangular wrapping. At higher density labeled III, the LLs have degeneracy 8 corresponds to two pockets. All these regions are separated by resistive ridges (region V) in the $n$-$D$ map, indicative of Lifshitz transitions where the topology of the Fermi surface abruptly changes[26,20,27].

These multiple Lifshitz transitions arise naturally from the calculated band structure of ABCBC graphene (Fig. 3c). As $n$ (or the chemical potential) is varied under positive or negative layer potentials, the Fermi surface geometry undergoes several abrupt changes. Our theoretical model predicts such transitions, matching the experimental observations of resistance peaks and LL changes. In short, by gating the system, we drive the Fermi level across van Hove singularities of the combined flat bands, yielding multiple topology changes in the Fermi sea.

**Low-field $|\nu| = 6$ quantum Hall state**

A particularly striking result is seen in the quantum Hall regime. At near zero $D$, we find that the first well-developed Hall plateau appears at filling factor $\nu = -6$ (minus sign represents hole side). Figure 4a shows the LL fan diagram at $D = 0$, where at $B =$ 26 mT, the $\nu = -6$ plateau is already visible as the dominant gap. Figure 4b plots the Hall resistance $R_{xy}$ and $R_{xx}$ as a function of $B$ at $n = -1\times10^{10}$ cm$^{-2}$. $R_{xy}$ rapidly approaches $h/6e^2$ (~93% quantization) by 26 mT, and $R_{xx}$ simultaneously drops to a minimum. This $\nu = -6$ state persists over a range of $D$ roughly from –0.06 to +0.05 V/nm, as shown in Fig. S8a. Furthermore, under slightly negative displacement fields ($D < -0.2$ V/nm), the $|\nu| = 6$ state continues to be the first developed LL on both electron and hole sides (Fig. 4c, Fig. S7&8).

There are two possible mechanisms for the observed $|\nu| = 6$ state. The first is a conventional Landau level (LL) quantization scenario. In this picture, the $|\nu| = 6$ plateau originates from the zero-energy LL structure associated with the ABC trilayer-like cubic band component. In pure ABC trilayer graphene, the $N = 0,1,2$ LL orbitals are degenerate at zero energy, yielding a 12-fold degenerate LL when accounting for spin and valley degrees of freedom[28]. Our band structure calculations (Fig. S10b) show that the flat valence band at $\Delta = 0$ is mainly composed of ABC trilayer states, supporting this interpretation. However, at $\Delta = -25$ meV, ABC and AB layers contribute almost equally in top valence band and bottom conduction band, which cannot explain the origin of $|\nu| = 6$ when $D < -0.2$ V/nm.

An alternative explanation is that the $|\nu| = 6$ state arises from a Chern insulator induced by spontaneous valley polarization under a small magnetic field. In this case, the internal built-in electric fields, originating from the non-centrosymmetric stacking, break inversion symmetry and lift the valley degeneracy without requiring Landau level quantization. The application of a weak magnetic field then favors one valley over the other, leading to a nonzero total Chern number. Such Chern insulator behavior has been proposed and observed in other systems, such as rhombohedral pentalayer graphene under $D$ in a small magnetic field[22,29,30]. Though we cannot distinguish between these two scenarios. In summary, the $|\nu| = 6$ quantum Hall state reflects the underlying flatband structure of ABCBC graphene and highlights the interplay between spontaneous symmetry breaking, Berry curvature effects, and magnetic field response in mixed-stacked multilayer graphene.

**Discussion and summary**

We note that, despite the intrinsic polarization of the ABCBC stacking, all our devices do not exhibit ferroelectric hysteresis under gating. This is likely because our samples contain a single stacking domain, without domain walls or a reversed-stacking seed (e.g. an ABABC region), the built-in dipole cannot be easily switched. In contrast, recent experiment on mixed-stacked tetralayer graphene has demonstrated switchable ferroelectric behavior when domain boundaries are present[31,32]. Thus, stacking-induced polarization is confirmed, but requires extra ingredients, such as domain wall induced stacking sliding, to show memory effects.

Although our transport measurements focus on devices identified as ABCBC-stacked pentalayer graphene, we note that another possible stacking configuration, ABCAC, shares the same non-centrosymmetric lattice symmetry and exhibits similar built-in layer polarization. In the chiral decomposition picture, ABCAC corresponds to a 1+4 partitioning (a monolayer coupled to a chiral tetralayer), in contrast to the 3+2 decomposition of ABCBC. To rule out ABCAC, two pieces of evidence are provided. Firstly, according to band calculation, in high-energy level, ABCBC harbors three isolated bands (labeled as $v_1$, $v_2$, $v_3$ in Fig. S2d), but ABCAC harbors two equal-energy bands crossing with each other (labeled as $u_2$, $u_3$ in Fig. S2e) and another band (labeled as $u_1$) shifted a little from them. The calculated high-energy band of ABCBC can well fit the measured NanoARPES dispersion (Fig. 1c and Fig. S2f). Secondly, we performed theoretical calculations of band gap under vertical displacement fields (Fig. S10b). We find that in ABCAC stacking, the intrinsic gap at charge neutrality vanishes rapidly for both positive and negative directions of applied displacement field. This behavior differs markedly from our experimental observations on ABCBC devices, where the gap exhibits a strong asymmetry and persists for negative fields. Thus, both NanoARPES dispersion and the field dependence of the gap provides clear evidence to rule out ABCAC as the stacking sequence in our devices.

It should be mentioned that M1 has ~12.8 nm moiré superlattice but the moiré effect is very weak (Fig. S9) and has negligible influence on most phenomena we observed.

In summary, we have performed the first transport study of ABCBC mixed-stacked pentalayer graphene. The observation of an intrinsic band gap at $D$ = 0 and its asymmetric tuning by the electric field are in quantitative agreement with the non-centrosymmetric lattice and built-in layer potentials. Furthermore, we uncover multiple Lifshitz transitions and a complex Landau level diagram arising from the interplay of the cubic and parabolic bands. The emergence of a $\nu$ = -6 quantum Hall state at extremely low magnetic field highlights the interplay between spontaneous symmetry breaking, Berry curvatures, and magnetic field response in mixed-stacked multilayer graphene. More broadly, our work highlights mixed-stacked multilayers as a rich platform of coexisting flat bands and tunable symmetries. In analogy to correlated states seen in moiré graphene systems[33–36,16,37–40], one may now explore similar phenomena (correlated insulators, unconventional superconductivity, magnetism, etc.) in these natural crystals without any twist, even in non-centrosymmetric lattices[14,22,41,42]. The ability to control layer polarization, band topology, and carrier density in a clean, gate-tunable setting opens new avenues for exotic quantum electronic phases and device applications.

## Methods

**Sample fabrications** Graphene, graphite and hBN are mechanically exfoliated on $SiO_2$(285nm)/Si substrates, and the layer numbers are identified using optical contrast and atomic force microscopy. The stacking order of pentalayer graphene is identified using IR CCD and SNOM. A dry transfer method using polycarbonate (PC) or polypropylene carbonate (PPC) is implemented to construct the heterostructures. For the sample measured by nanoARPES, hBN and ABCBC were sequentially picked up by PPC film. The ABCBC/BN was then flipped and transferred to a clean $SiO_2$/Si wafer, which was then annealed at 350°C in vacuum to remove the PPC film underneath the heterostructure. Standard e-beam lithography, reactive ion etching and metal evaporation are conducted to make the devices into Hall bar geometry with the one-dimensional edge contacts. After each step of transfer and fabrications, SNOM imaging is performed to check the stacking orders of graphene.

**Transport measurements** Most of transport measurement is done in the 1.5 K base temperature Oxford variable temperature insert (VTI) system. Some other transport measurement is done in 2 K base temperature system (Electronics Transport Measurement System, Model EM7, East Changing Technologies, China). The measurement below 1.5 K is conducted in a top-loading dilution refrigerator (Oxford TLM), in which the sample is immersed in the 3He-4He mixture during the measurements. Stanford research system SR830, SR860, NF LI5650 and Guangzhou Sine Scientific Instrument OE1201 lock-in amplifiers with an alternating-current of 10 ~ 500 nA at a frequency of 17.7 Hz in combination with a 10 MΩ resistor are used to measure the resistance. Keithley 2400 source meter is used to apply the gate voltages. The displacement field $D$ is set by $D = (D_b+D_t)/2$, and carrier density is determined by $n = (D_b-D_t)/e$. Here, $D_b = +\varepsilon_b(V_b - V_b^0)/d_b$, $D_t = -\varepsilon_t(V_t-V_t^0)/d_t$, where $\varepsilon$ and $d$ are the dielectric constant and thickness of the dielectric layers, respectively, $V_b^0$ and $V_t^0$ are effective offset voltages caused by environment-induced carrier doping.

**ARPES measurements** Before NanoARPES measurements, the sample was annealed at 180 °C for several hours in ultrahigh vacuum (UHV). NanoARPES measurements were performed at beamline ANTARES of the Synchrotron SOLEIL in France, using a photon energy of 95 eV and linear horizontal (LH) polarisation, with a beam size smaller than 1 μm. The sample was measured at 80 K in a vacuum better than $3\times10^{-10}$ mbar, with the overall energy resolution of 40 meV.

**Ab initio calculations** We employed the Vienna ab initio simulation package (VASP)[43] to simulate electronic properties of ABCBC-stacked pentalayer graphene in the framework of density functional theory (DFT)[44,45]. The electrons are described with the Perdew-Burke–Ernzerhof functional (PBE)[46] in the generalized gradient approximation. A cutoff energy of 600 eV and a k-mesh of 36 × 36 × 1 are adopted. The graphene lattice constant and the interlayer distances are set to $a$ = 2.46 Å and $c$ = 3.35 Å. The maximally-localized Wannier functions[47] were generated using $C$-$p_z$ orbital. With the tight-binding Hamiltonian constructed by the WANNIER90 package[48], the band gap versus electrical field were calculated using the WANNIERTOOLS software package[49].

**SWMcC model** After fitting the band structure from Ab initio calculations, we get the hopping terms for the Sloncewski Weiss-McClure(SWMcC) model[50–52] to construct effective model at $K_{\pm}$ points (Fig. S10). In the basis of A, B sublattice of ABCBC, the Hamiltonian is

$$H=\begin{pmatrix} H_0+\frac{1}{2}(1+\sigma_z)\Delta'+\Delta_{hBN} & V_{AB} & W_{ABC} & 0 & 0 \\ V_{AB}^{\dagger} & H_0+\Delta' & V_{AB} & W_{ABA} & 0 \\ W_{ABC}^{\dagger} & V_{AB}^{\dagger} & H_0+\Delta'(1-\sigma_z) & V_{AB}^{\dagger} & W_{BAB} \\ 0 & W_{ABA}^{\dagger} & V_{AB} & H_0+\Delta'(1+\sigma_z) & V_{AB} \\ 0 & 0 & W_{BAB}^{\dagger} & V_{AB}^{\dagger} & H_0+\frac{1}{2}(1-\sigma_z)\Delta'+\Delta_{hBN} \end{pmatrix}$$

where $H_0=\begin{pmatrix} 0 & t_0\pi^{\dagger} \\ t_0\pi & 0 \end{pmatrix}$ , $V_{AB}=\begin{pmatrix} -t_4\pi^{\dagger} & t_3\pi \\ \gamma_1 & -t_4\pi^{\dagger} \end{pmatrix}$ , $W_{ABA}=\begin{pmatrix} \gamma_2/2 & 0 \\ 0 & \gamma_5/2 \end{pmatrix}$ ,

$W_{BAB}=\begin{pmatrix} \gamma_5/2 & 0 \\ 0 & \gamma_2/2 \end{pmatrix}$, $\sigma_z=\begin{pmatrix} 1 & 0 \\ 0 & -1 \end{pmatrix}$, $\pi=(\zeta p_x+ip_y)\sqrt{3}a/2\hbar$, $\zeta=\pm 1$ is the valley index.

Hopping terms and onsite terms are shown in Table 1[53]. The onsite energy $\Delta'$ originates from the dimmer bonds between adjacent layers, the factor by which $\Delta'$ is multiplied depends on the number of interlayer couplings to that site. $\Delta_{hBN}$ is an additional on-site potential induced by substrate as 18 meV[54].

To add electric field $E$, we set the third layer as the zero potential point, and the onsite energy differences induced by electric field is $\Delta = eEd$, where $d = 3.35$ Å is the distance between two nearby layers. To better satisfy the experiment, we shift $\Delta$ by 5 meV.

**Acknowledgments**

G.C. acknowledges support from NSF of China (grant nos. 12350005 and 12174248), National Key Research Program of China (grant nos. 2021YFA1400100 and 2020YFA0309000), Shanghai Science and Technology Innovation Action Plan (grant no. 24LZ1401100), and Yangyang Development Fund. S.Z. acknowledges support from New Cornerstone Science Foundation through the XPLORER PRIZE. K.L. acknowledges support from the National Natural Science Foundation of China (grant no. 124B2071). K.L. and Y.S. acknowledge support from T.D. Lee scholarship. M.T. and N.W. acknowledge support from Open Project of Key Laboratory of Artificial Structures and Quantum Control (Ministry of Education), Shanghai Jiao Tong University. K.W. and T.T. acknowledge support from the JSPS KAKENHI (Grant Numbers 20H00354, 21H05233 and 23H02052) and World Premier International Research Center Initiative (WPI), MEXT, Japan. A portion of this work was carried out at the Synergetic Extreme Condition User Facility (SECUF, https://cstr.cn/31123.02.SECUF). We acknowledge SOLEIL for the provision of synchrotron radiation facilities of beamline ANTARES.

**Competing interests**

The authors declare no competing financial interests.

**Data availability**

All data that support the findings of this study are available from the contact author upon request.

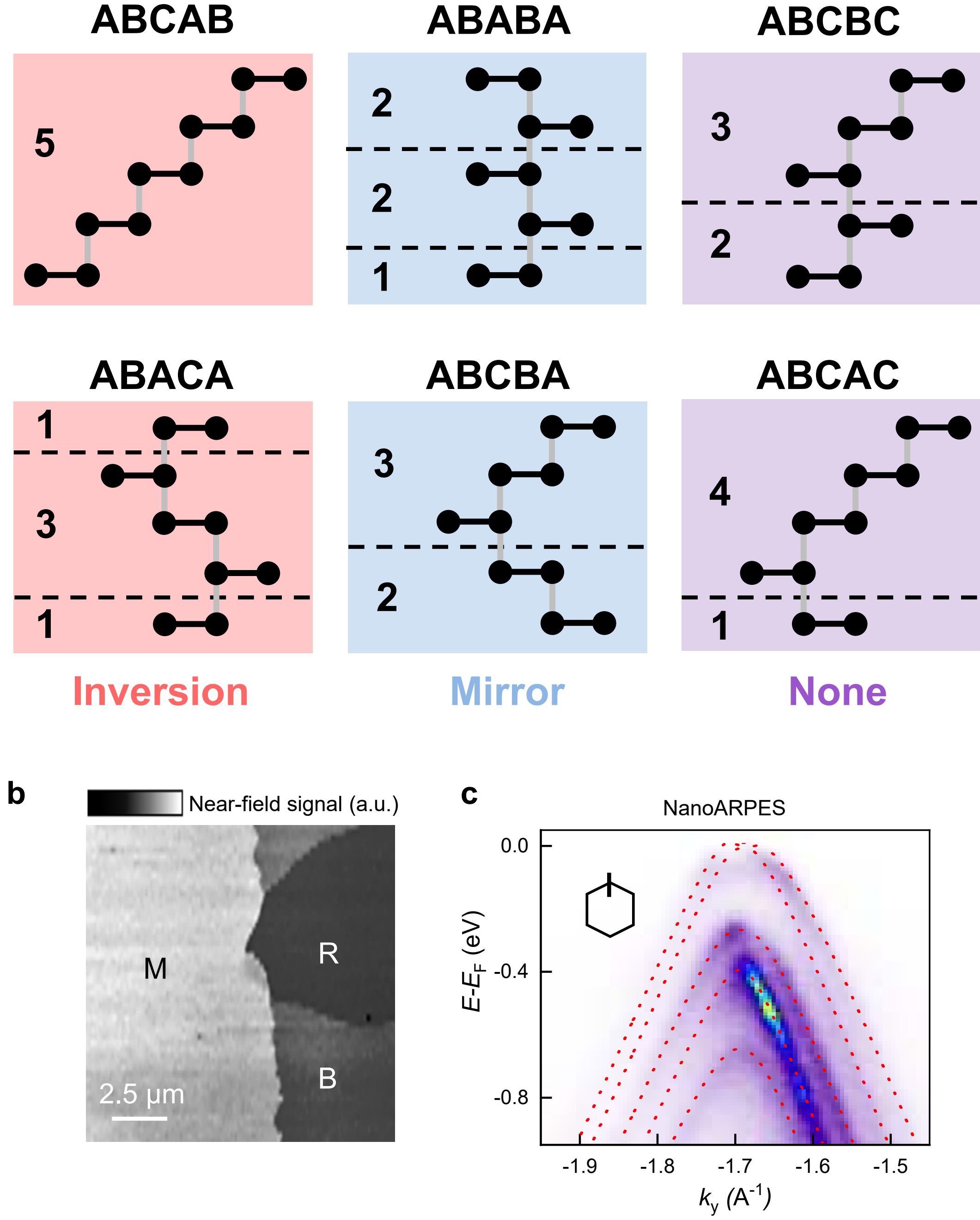

**Figure 1. Possible stacking orders of pentalayer graphene and their chiral decomposition.** **a,** Lattice structures of all six possible stacking orders of pentalayer graphene. The stackings are grouped by inversion, mirror and non-centrosymmetry. The number labeled around the lattice is the chiral decomposition result of each stacking. For example, ABABA is labeled as 2+2+1, it means the low energy bands are composed of three parts, which are two parabolic bands ($k^2$) and one linear band ($k^1$). **b,** Near-field infrared nanoscopy images of pentalayer graphene exfoliated on $SiO_2$/Si substrate. The wavelength of incident light is 10.6 μm. **c,** Electronic band structure of ABCBC attained from NanoARPES measurements and single-particle calculation (red dotted lines). The dispersion image measured parallel to Γ-K direction (vertical line in the inset)

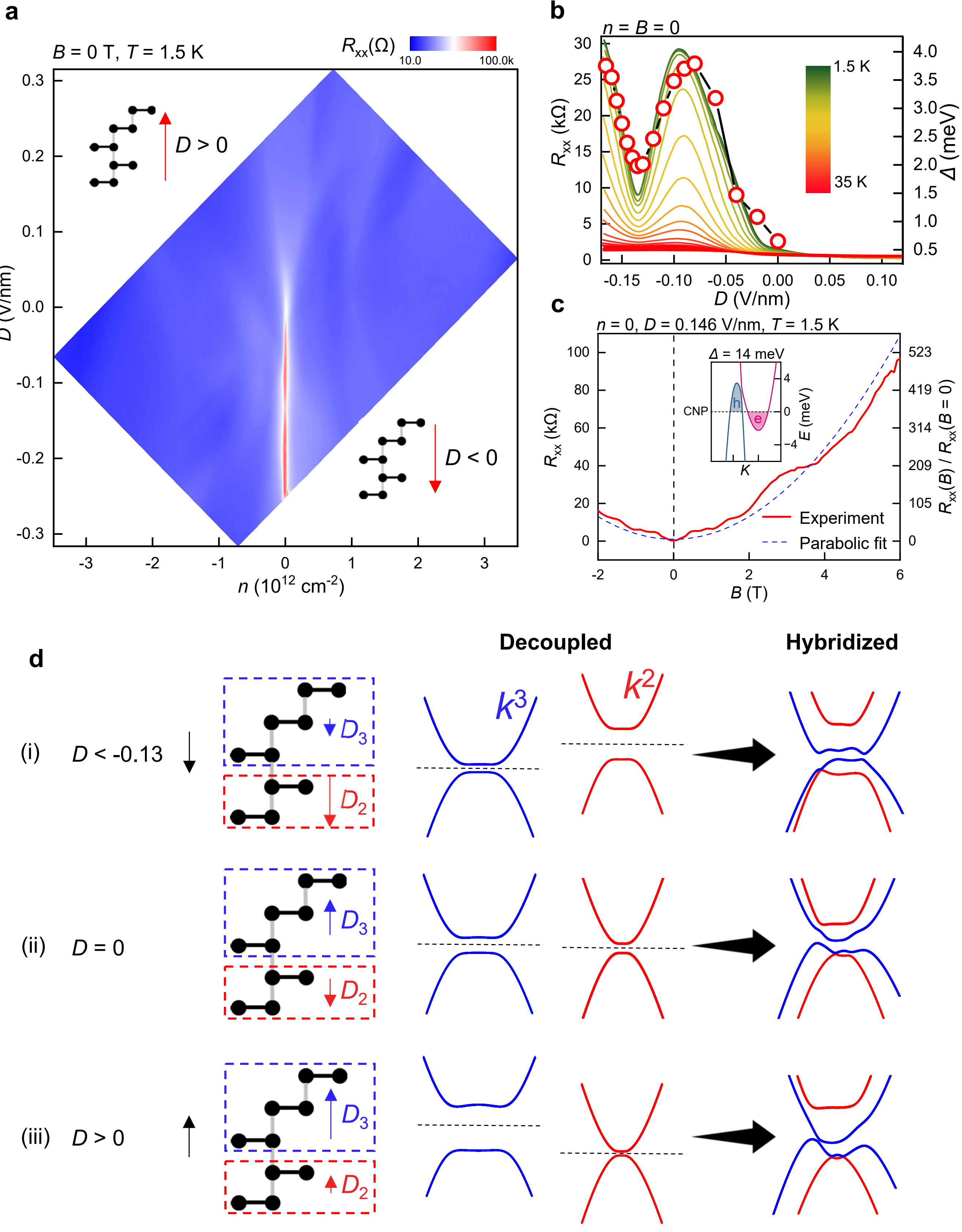

**Figure 2. Symmetry breaking and layer polarization of ABCBC. a,** Color plot of resistance $R_{xx}$ as a function of carrier density $n$ and displacement field $D$. The color bar is in the log scale. **b,** $D$-dependent $R_{xx}$ for $n = 0$ at varied temperatures from 1.5 K to 35 K, the red hollow circles are the corresponding transport gap extracted from Arrhenius plot. **c,** Magnetoresistance as a function of magnetic field. The left axis is the longitudinal resistance and the right axis is relative magnetoresistance defined as $R(B)/R(0)$, where $R(0)$ is the resistance at $B = 0$. The inset is the zoom-in band structure around Fermi surface at $+|D|$ side. **d,** Illustrations of the layer polarization and the band gap evolution at different $D$.

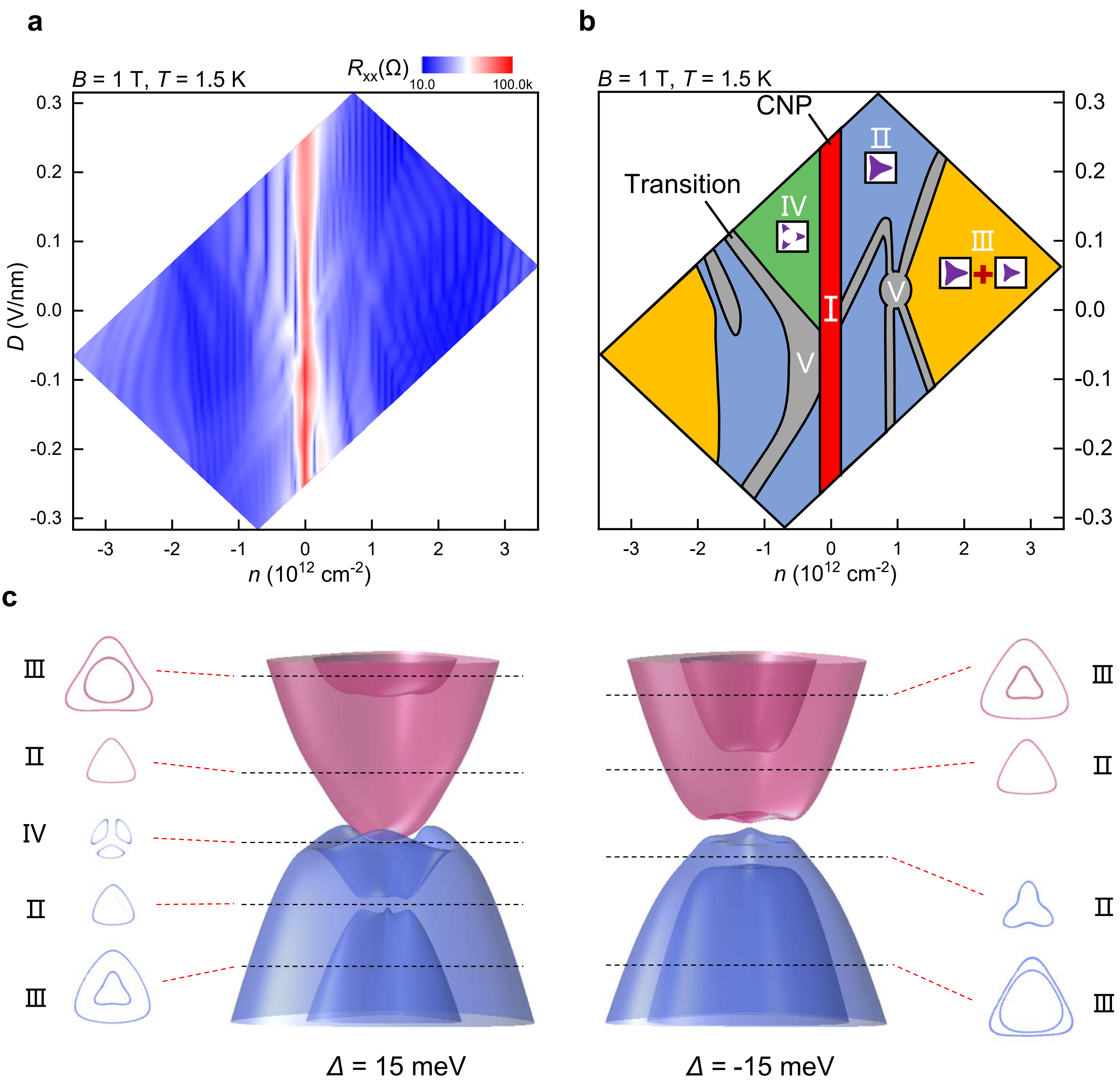

**Figure 3. Multiple Lifshitz transitions under *D*. a,** *n* - *D* color plot of resistance at *B* = 1 T. The color bar is in the log scale. **b,** Corresponding phase diagram of **a** according to the Landau level degeneracy in each region. **c,** Calculated single particle band structures and the Fermi surfaces at different Fermi energy. Conduction and valence bands are drawn as red and blue respectively.

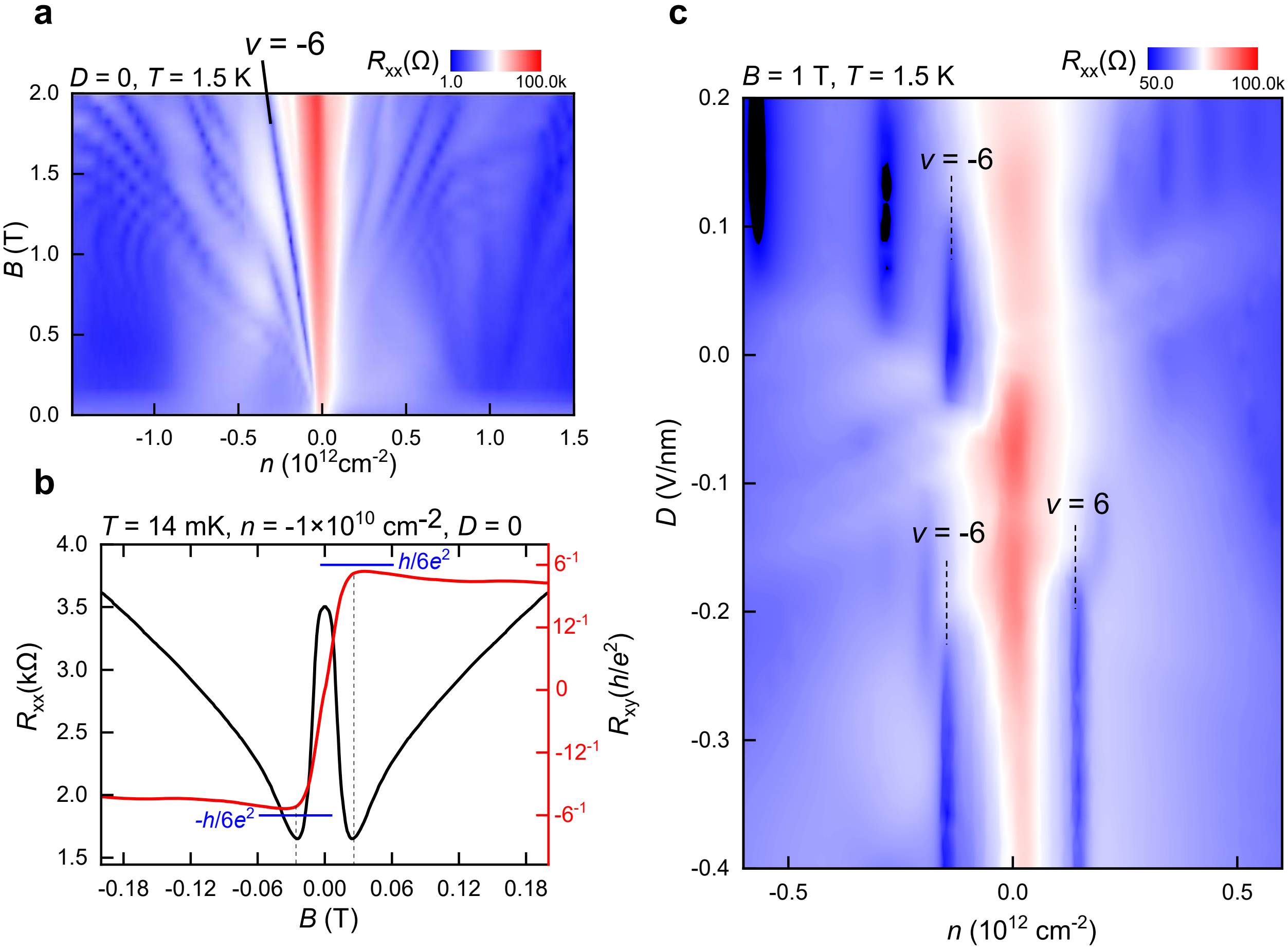

**Figure 4. Robust |ν| = 6 quantum hall state. a,** $R-n-B$ color plot at $D=0$ when $T=1.5$ K for device M1. **b,** Corresponding Hall resistance $R_{xy}$ and $R_{xx}$ as a function of magnetic field measured at $D=0$, $n=-1\times10^{10}$ cm$^{-2}$, $T=14$ mK. **c,** $R-n-D$ color plot at $B=1$ T when $T=1.5$ K for device M2.

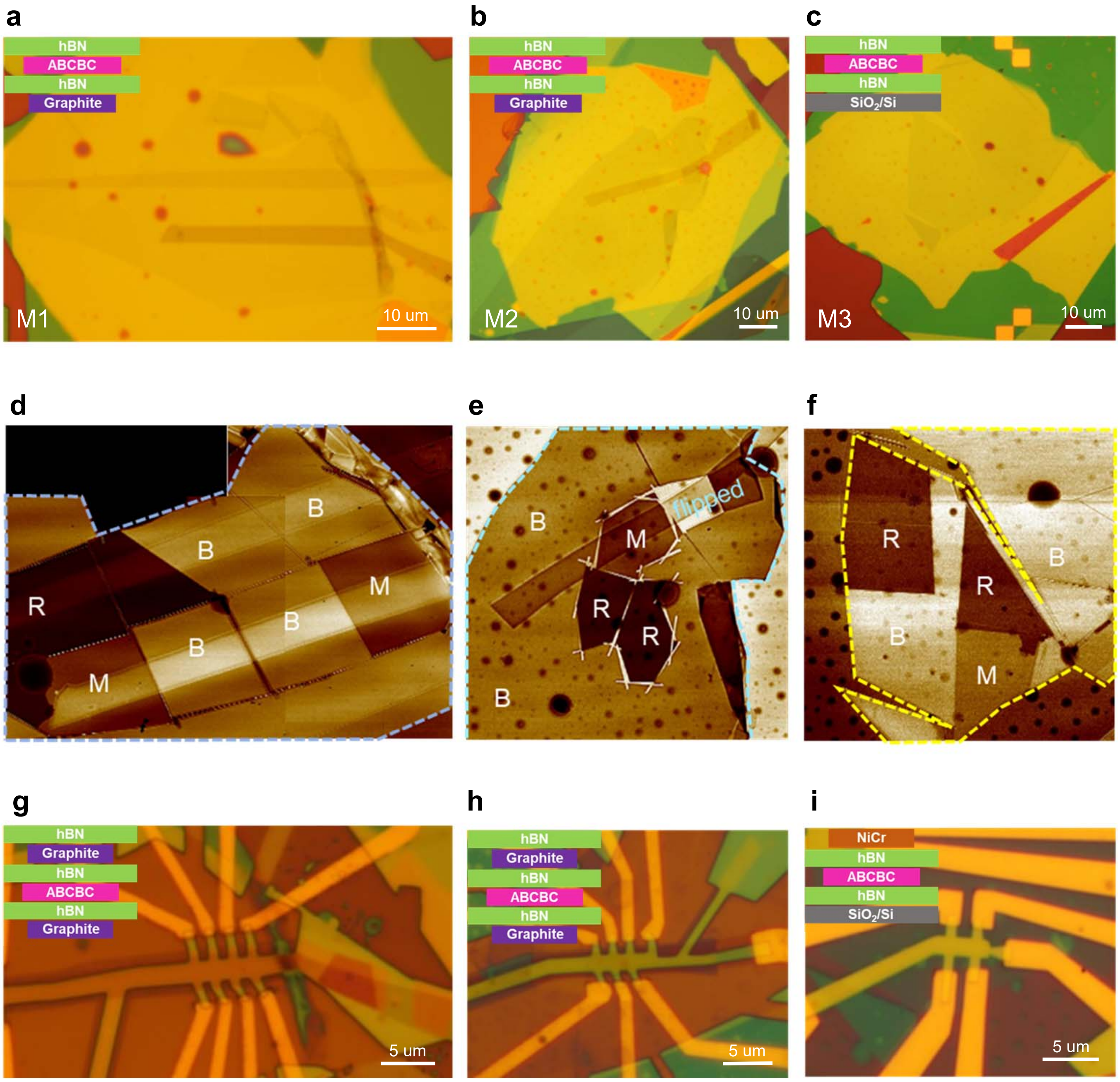

**Figure S1. Identification of ABCBC. a~c,** Optical images of samples M1, M2, M3 before transferring graphite or evaporating metal top gate onto. **d~f,** Corresponding SNOM images with incident light wavelength 6.3 μm. The darkest region is ABCBA, the brightest region is ABABA, ABCBC shows middle contrast. This is different with SNOM images of graphene flakes on SiO2/Si substrate shown in Fig 1b, in which M-stacking domain shows the brightest contrast, since the contrast can be tuned by the interference effect of different substrates. **g~i,** Optical images of M1, M2 and M3 after being made into Hall bar devices.

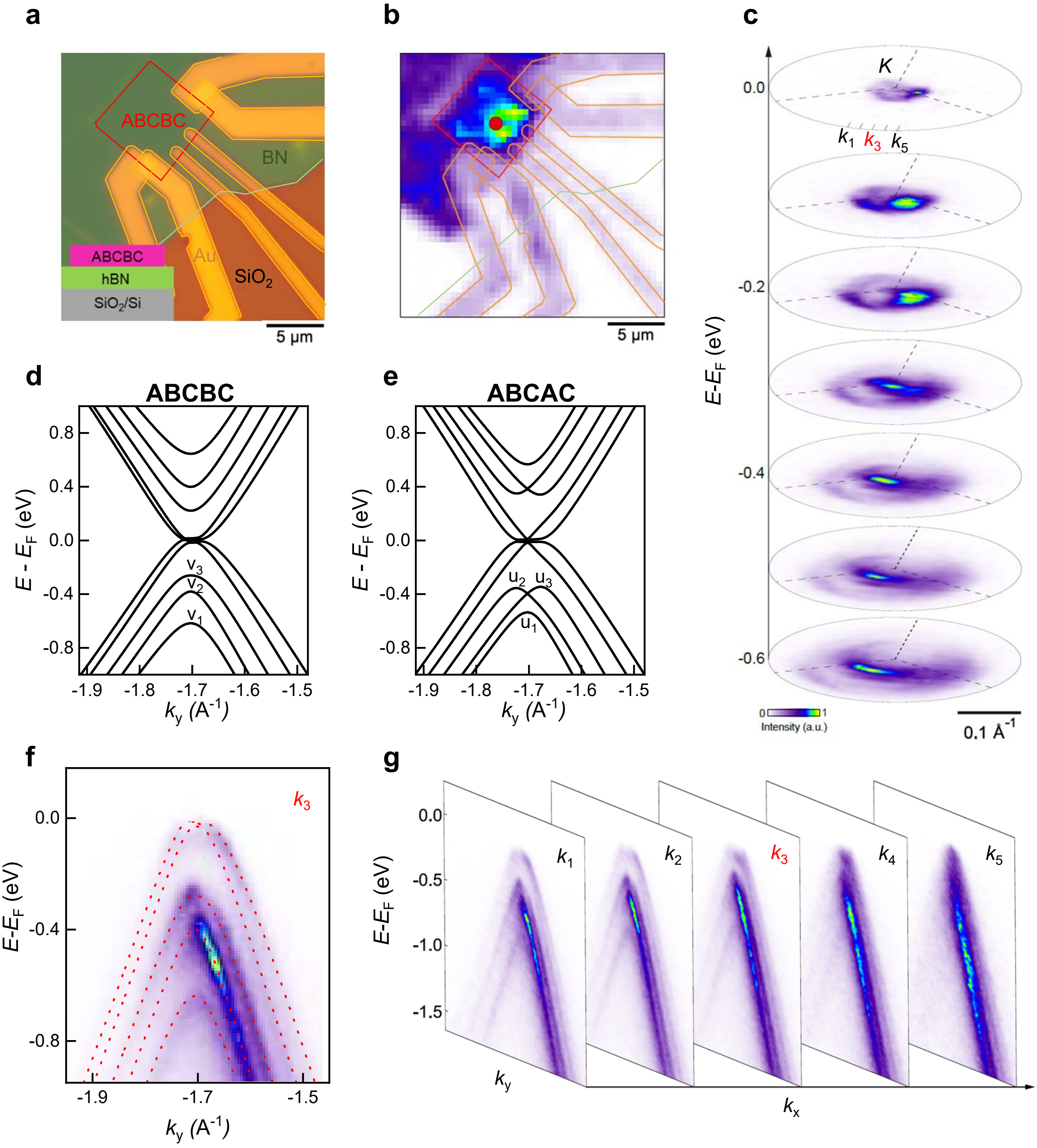

**Figure S2. NanoARPES measurement of device M4. a,** Optical image of ABCBC/hBN on silicon wafer, where ABCBC, Au electrodes and hBN are marked by red, orange and green curves. **b,** NanoARPES spatial image measured at the same area as the optical image. **c,** Intensity maps measured at energies from Fermi energy ($E_F$) to -0.6 eV. **d** and **e,** Calculated single particle band structures of ABCBC and ABCAC, respectively. **f** and **g,** Dispersion images of ABCBC measured along momentum cuts near the $K$ point, as indicated by black and red lines in **c**. Red dotted lines in **f** are the calculated band structure which can fit the experimental results well.

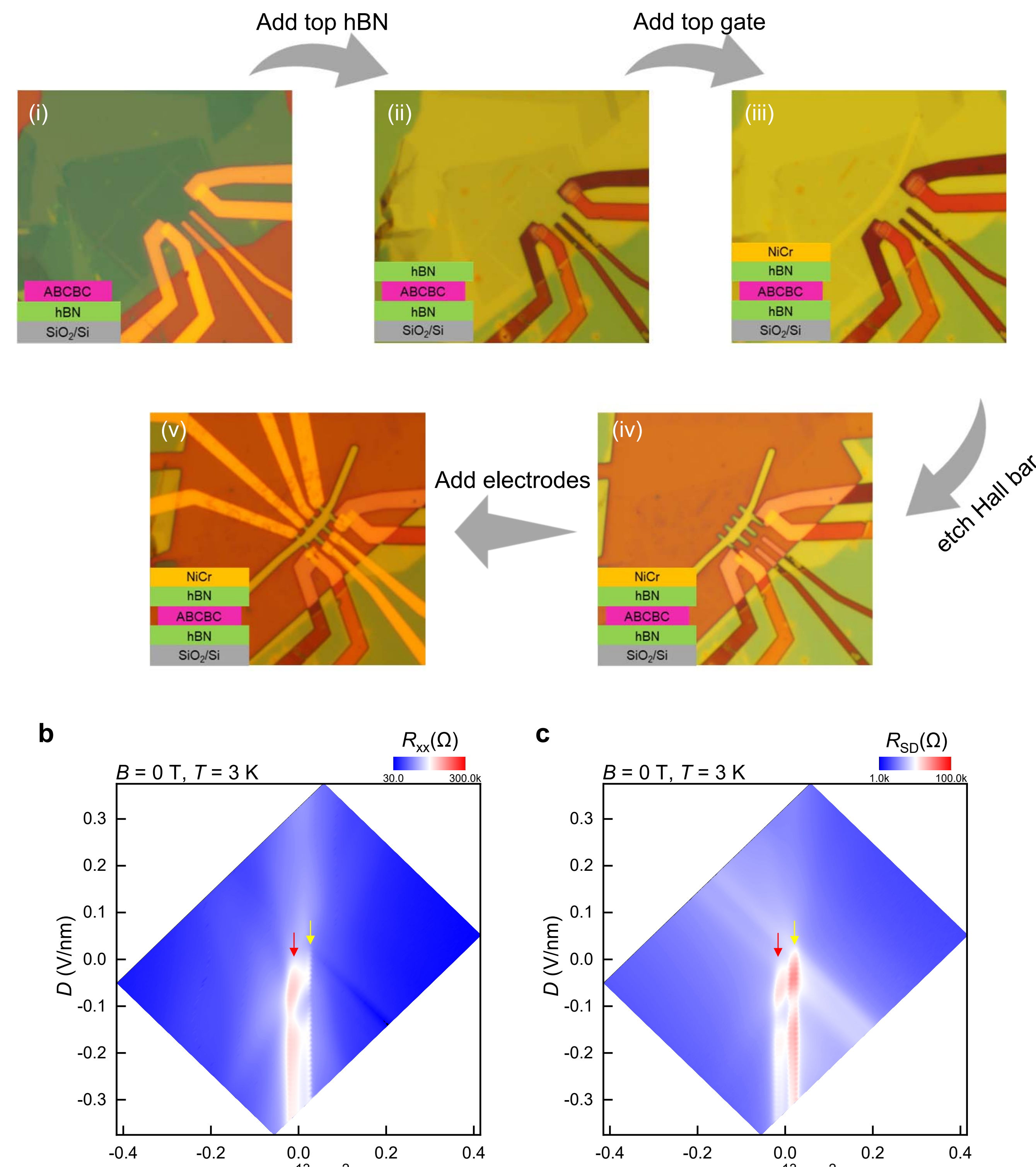

**Figure S3. Transport measurement of M4. a,** Workflow for transfer and nanofabrication process of the metallic gate/hBN/ABCBC/hBN/silicon gate device. The sample in (i) is the same with that shown in Fig. S2, another hBN with thickness ~30nm was transferred on top of it (ii), followed by e-beam deposition of nichrome about 12nm (iii). Standard reactive ion etching (RIE) using $CHF_3$ and $O_2$ was conducted to etch the sample into Hall bar (iv). Finally, electrodes were deposited with Au(50nm)/Cr(5nm) (v). **b** and **c,** n-D color plots of four- and two-terminal longitudinal resistance for M4, respectively. At CNP, there shows nonmonotonic $D$ dependence similarly with that in Fig. 2a. It should be mentioned that in both color plots there show two peaks, marked by red and yellow arrows in **b** and **c**, it suggests there are two domains with different intrinsic doping along source and drain.

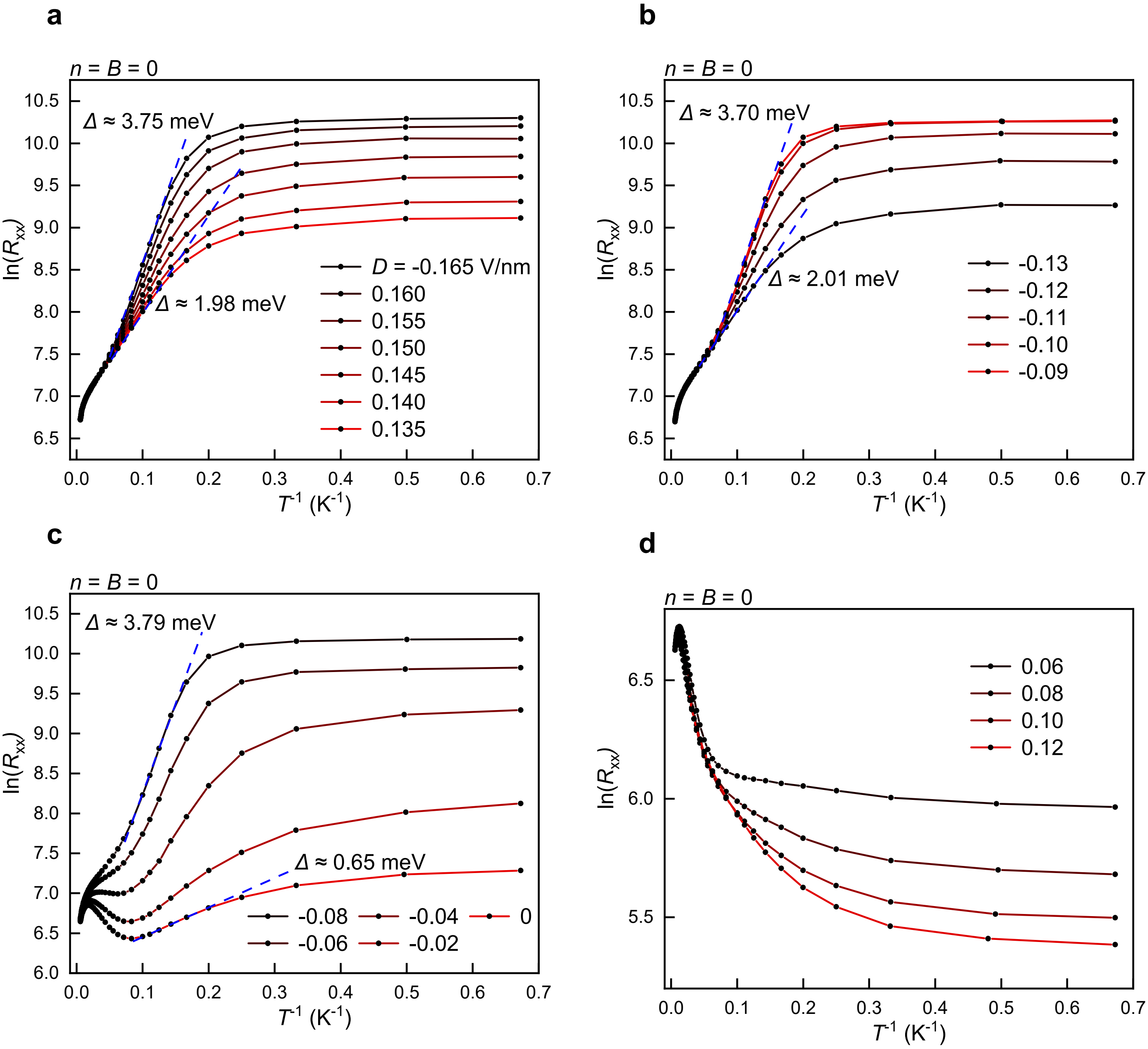

**Figure S4. Arrhenius plot for M1. a~d, .** $\ln(R_{xx})$ versus temperature $T$ at different displacement fields $D$. The transport gap $\Delta$ in Fig. 2b is extracted according to thermal activation equation $R_{xx} \propto e^{-\Delta/2k_{B}T}$. The dashed lines in **a** to **c** represent the linear regions to fit the gap. The sample shows a metallic behavior at low temperature in **d**.

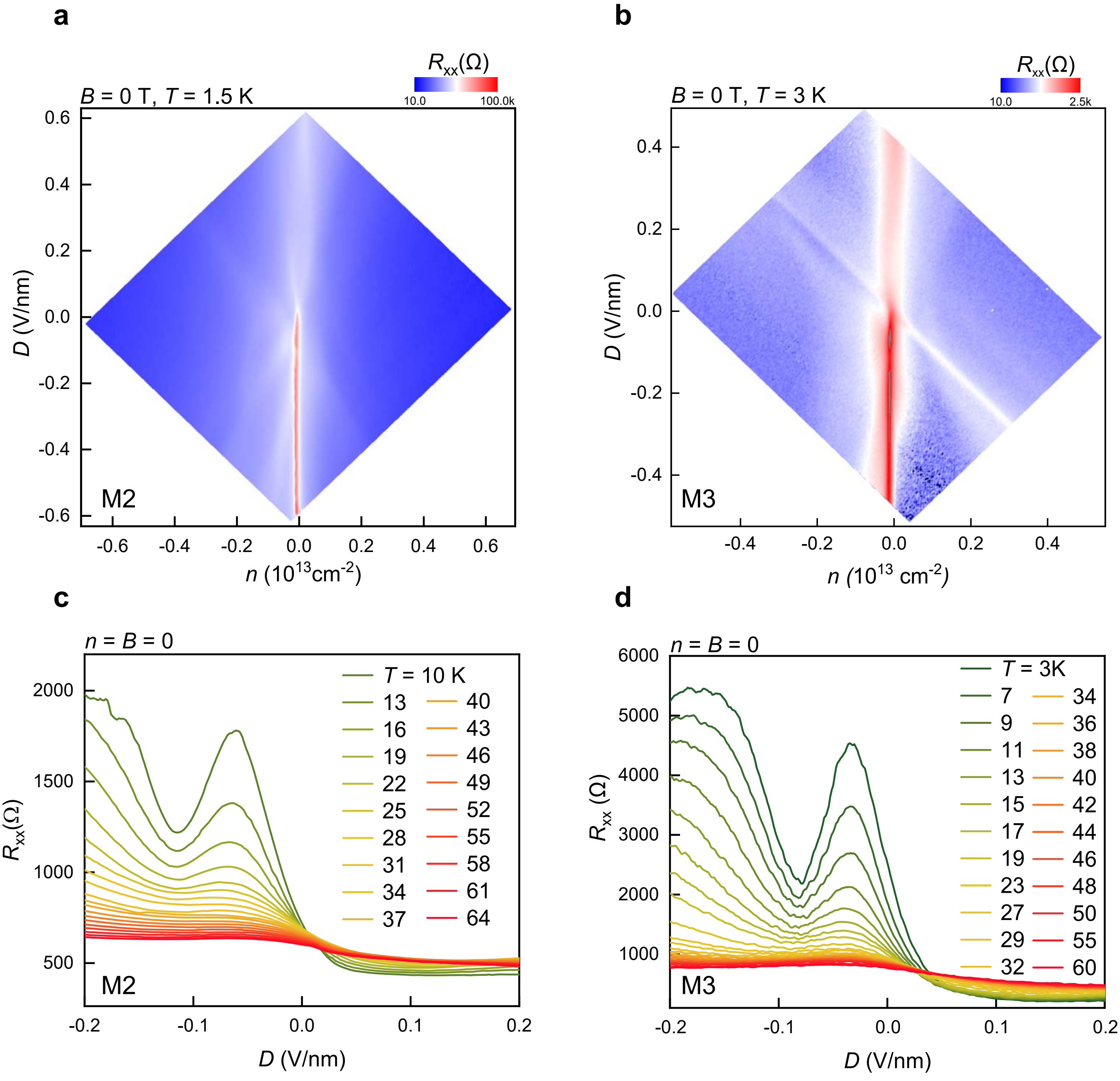

**Figure S5. Centrosymmetry breaking in M2 and M3. a&b,** Color plots of resistance $R_{xx}$ as a function of carrier density $n$ and displacement field $D$ for device M2 and M3 at $B = 0$. The color bar is in the log scale. **c&d,** Plots of $R_{xx}$ at different $D$ and $T$ with fixed $n = B = 0$ for M2 and M3. Both devices exhibit an asymmetric dependence on $D$, consistent with the behavior of M1 discussed in the main text.

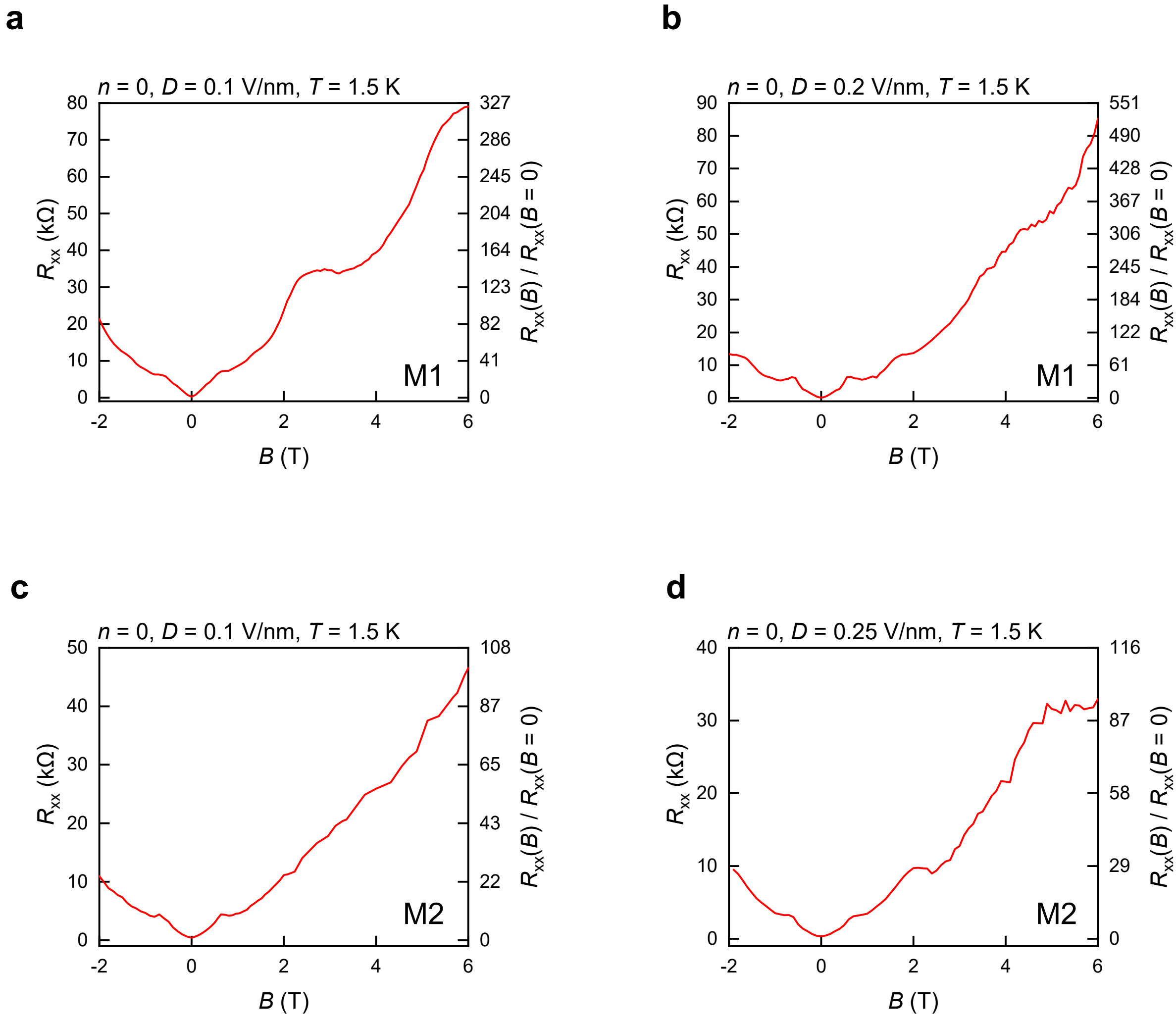

**Figure S6. Magnetoresistance in M1 and M2. a~d,** resistance as a function of magnetic field at different $+|D|$ at 1.5 K for M1 and M2. The smaller relative magnetoresistance in M2 is attributed to its lower sample quality compared to M1.

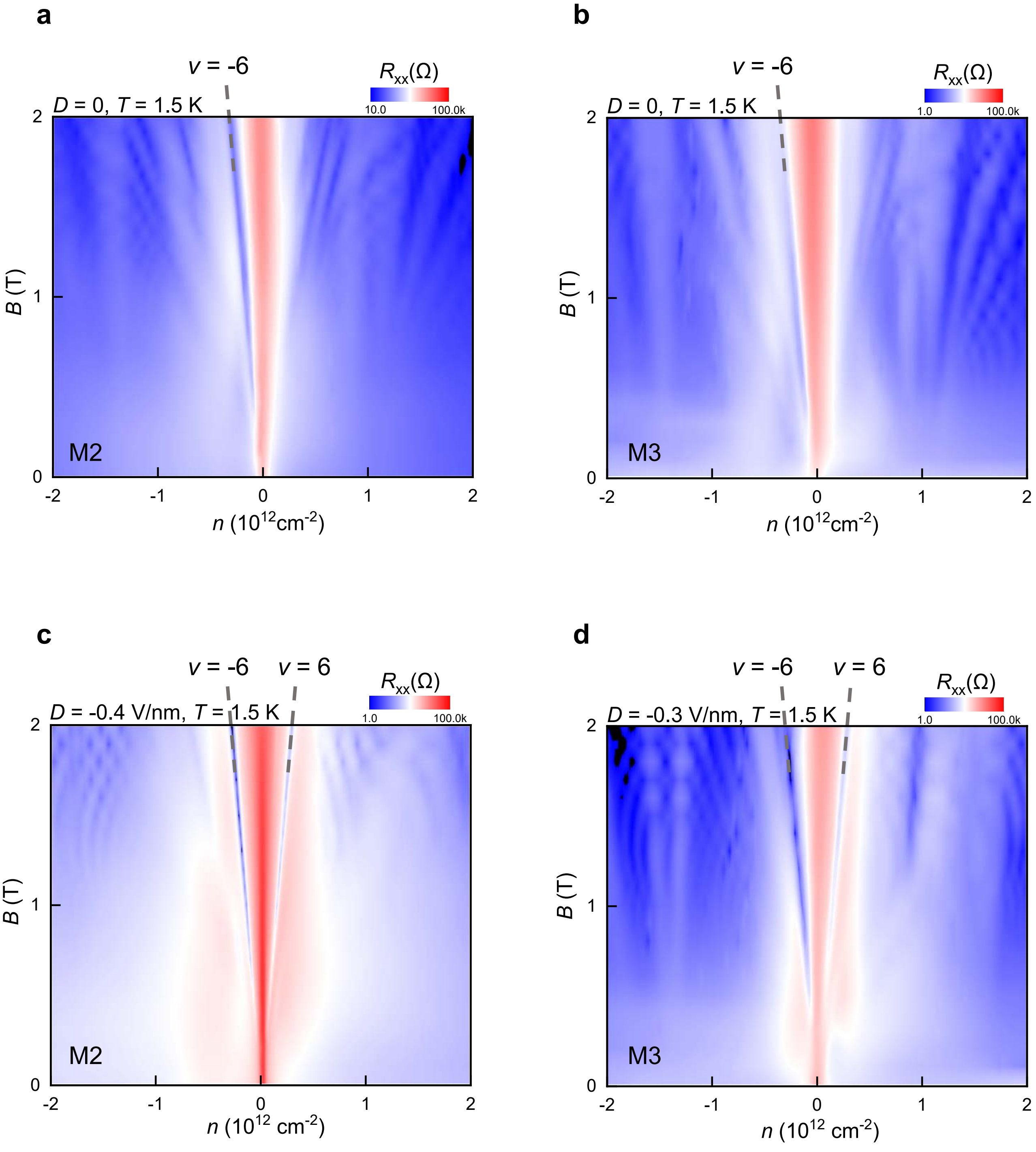

**Figure S7. Landau level fans for M2 and M3. a~d,** $R - n - B$ color plot of M2 and M3 at different $D$ when $T = 1.5$ K. As confirmed by the Streda formula, the $|\nu| = 6$ state is the first developed landau level in both devices.

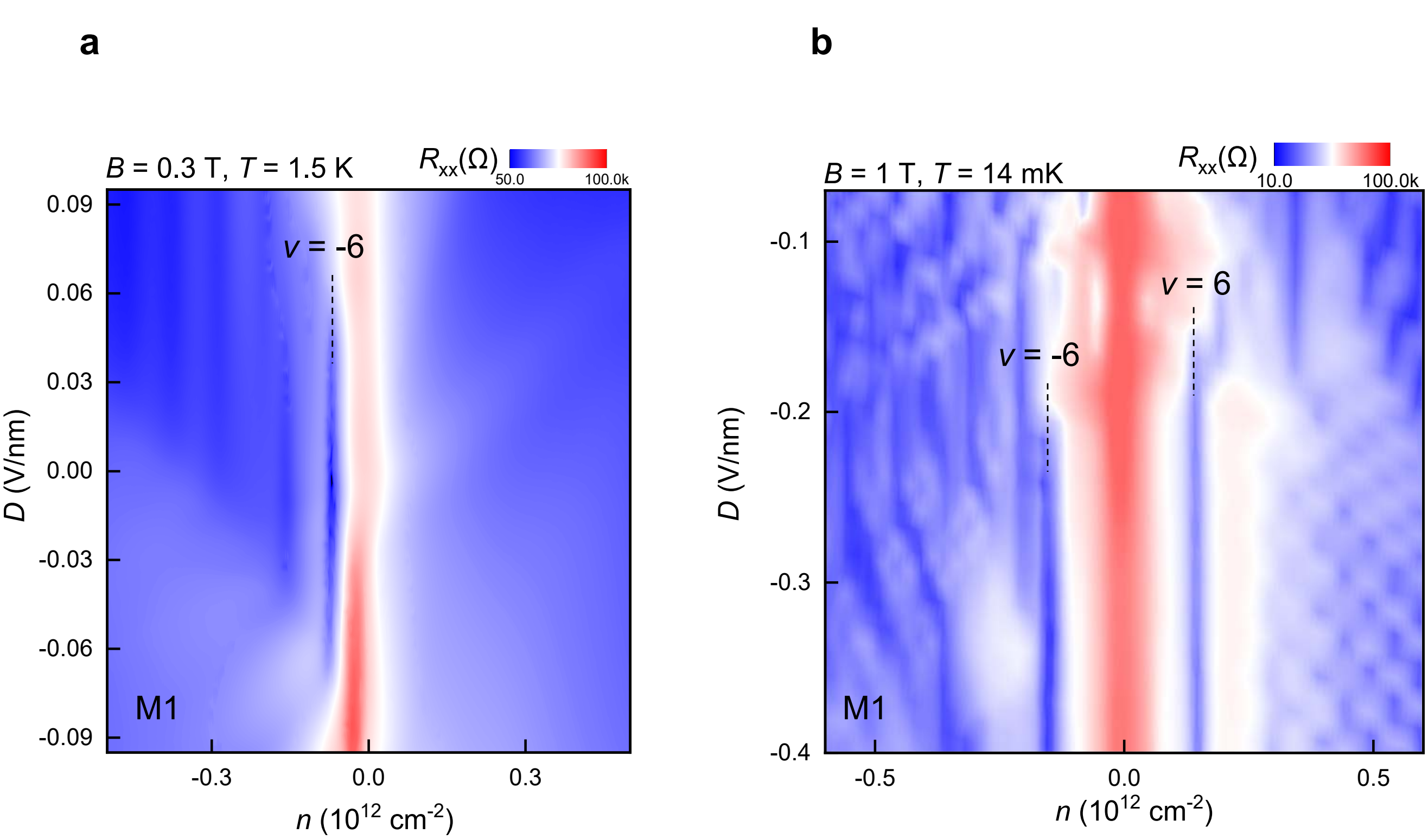

**Figure S8. $|\nu| = 6$ at certain $D$ in M1,** $R_{xx} - n - D$ color plot of M1 at $B$ = 0.3 T when $T$ = 1.5 K (**a**) and $B$ = 1 T when $T$ = 14 mK (**b**) respectively.

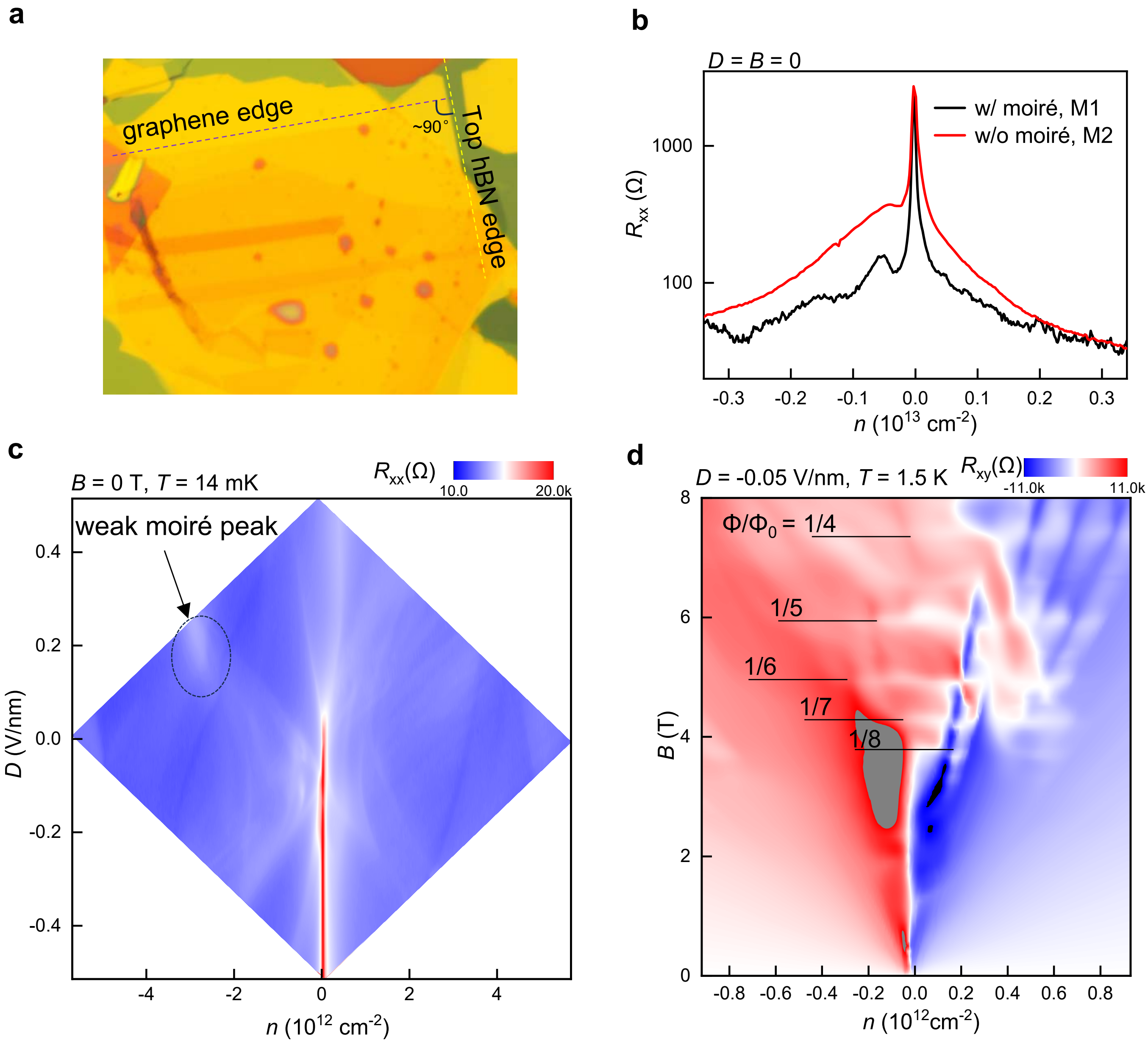

**Figure S9. Moiré effect in M1, a,** optical image of M1, the edges of graphene and top hBN are almost perpendicular to each other. **b,** $R_{xx}$-$n$ plot when $D = B = 0$, the behavior between M1 and M2 is similar shows no much difference. **c,** $R_{xx}$ - $n$ - $D$ color plot for M1 when $B = 0$, $T$ =14 mk. The moiré peak indicated by the black dashed circle is weak (smaller than 150Ω) and only appears when $D > 0.1$ V/nm and $n \approx -3\times10^{12}$ cm$^{-2}$. **d,** $R_{xy}$- $n$ - $B$ color plot when $D =$ -0.05 V/nm, $T = 1.5$ K. Brown-Zak oscillation can be seen and labeled by the black solid lines. According to the oscillation, 12.8 nm moiré wavelength is calculated.

Hopping terms (eV)

| $t_0$ | $t_3$ | $t_4$ | $\gamma_1$ | $\gamma_2$ | $\gamma_5$ |
|---|---|---|---|---|---|
| -3 | 0.3 | 0.066 | 0.39 | -0.017 | 0.038 |

Onsite terms (eV)

| $d_1$ | $d_2$ | $d_3$ | $d_4$ | $d_5$ | $d_6$ | $d_7$ | $d_8$ | $d_9$ | $d_{10}$ |
|---|---|---|---|---|---|---|---|---|---|
| 0.008 | 0.033 | 0.045 | -0.005 | 0 | 0.05 | 0.03 | 0.03 | 0.053 | 0.028 |

**Table 1**

**Table 1. SWMcC model,** a list of all the hopping and onsite terms used in the band calculation. Band structures shown in the main text are calculated according to the above parameters.

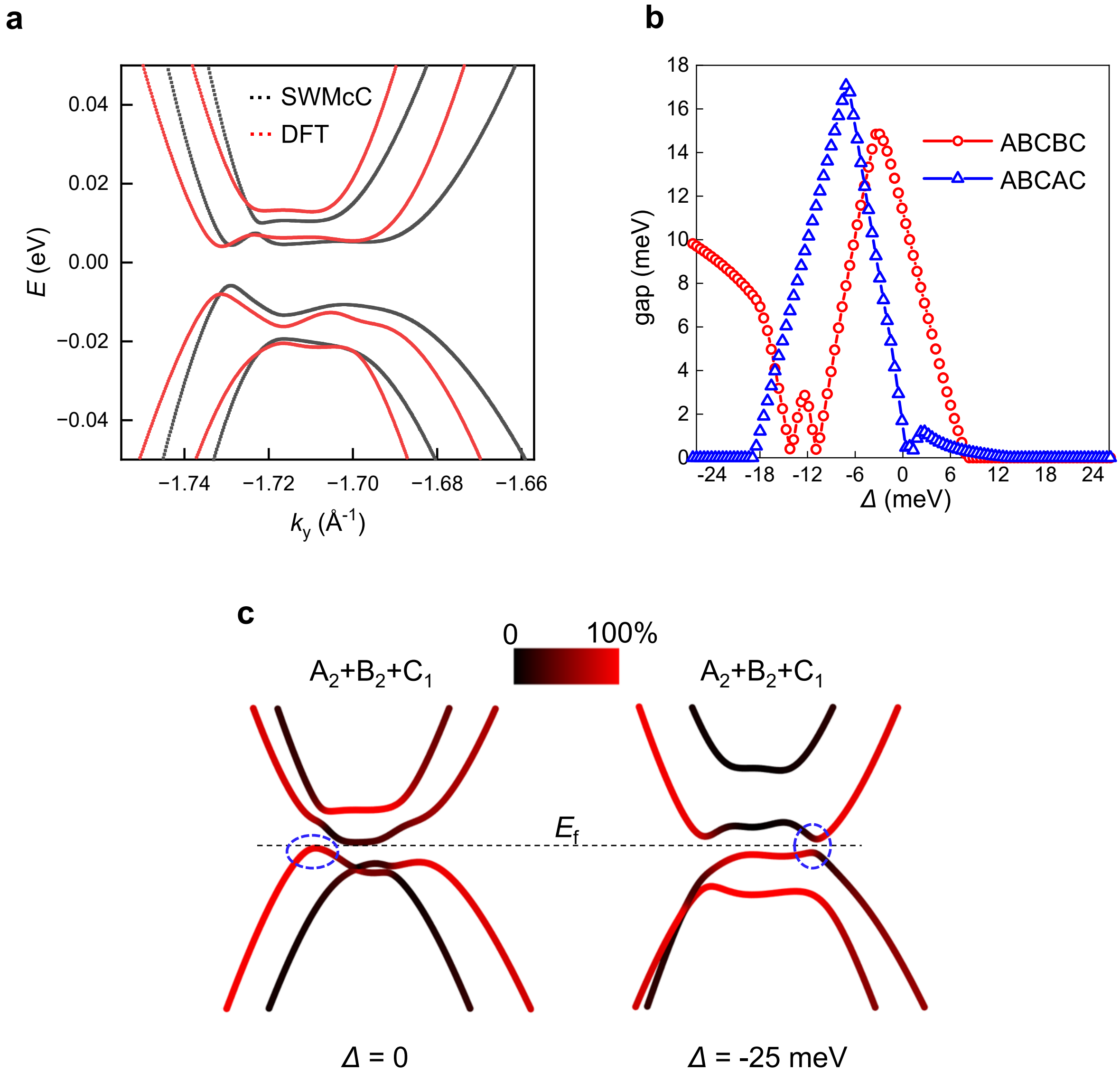

**Figure S10. Single particle band calculation. a,** Comparison between DFT and SWMcC model when $\Delta$ = -5 meV. **b,** Band gap with dependence of interlayer potential difference $\Delta$ according to tight-binding Hamiltonians obtained from maximally-localized Wannier functions and DFT. **c,** Occupation of ABC layers in the low energy band of ABCBC when interlayer potential difference $\Delta$ = 0 and -20 meV. If the color is red, the occupation is high, if it is black, the occupation is small. It is obvious the top part of valence band is almost occupied by ABC layer when $D$ = 0. However, for $\Delta$ = -25 meV, the occupation of ABC and AB layers are almost the same both in top valence band and bottom conduction band.